\documentclass{article}
\usepackage{amsmath}
\usepackage{mathrsfs}
\usepackage{mathtools}
\usepackage{xcolor}
\usepackage[margin=1in]{geometry} 
\usepackage{authblk}
\usepackage{fontawesome}
\usepackage{array} 
\usepackage{booktabs} 
\usepackage{amssymb}
\usepackage{geometry}
\usepackage{changepage}
\usepackage{ragged2e}
\usepackage{cancel}
\usepackage{mathrsfs}

\usepackage[T1]{fontenc} 
\usepackage[utf8]{inputenc} 
\usepackage{csquotes} 

\usepackage[numbers]{natbib}
\title{On the Natural Equivalence Between Canonical and Hilbert Energy–Momentum Tensors via Noether’s Theorem}
\author[1,\faEnvelopeO]{Peir-Ru Wang}

\affil[1]{Department of Physics, National Tsing Hua University, 30013 Hsinchu, Taiwan}
\affil[\faEnvelopeO ]{louisnthu@gapp.nthu.edu.tw}
\date{\today}

\renewcommand{\thesection}{\Roman{section}}
\begin{document}

\maketitle

\begin{abstract}
In this work, we investigate the structure and properties of the canonical energy-momentum tensor (EMT) across a range of field theories. We begin by developing a unified and systematic method that naturally yields the canonical EMT for gauge theory, without the need for artificial symmetrization or improvement terms. Our analysis highlights how Noether's theorem intrinsically emphasizes the 1-form nature of gauge potentials. We further extend to general relativity and demonstrate that the assumption of metric compatibility naturally implies a torsion-free connection. Building upon variational symmetry principles, we establish the equivalence between the Einstein–Hilbert EMT and the canonical EMT, thereby clarifying their respective roles in field dynamics and conservation laws. Lastly, we present an alternative derivation of the Einstein field equations via Noether's theorem.
\end{abstract}

\section{Introduction}

\noindent
The energy–momentum tensor (EMT) is a central object in theoretical physics, playing a dual role: it serves as the source term in Einstein’s field equations, and it appears as the conserved Noether current\cite{noether1971invariant} associated with spacetime translations. Its construction and properties are thus critical to understanding both field dynamics and the geometric structure of spacetime.

There are two widely accepted methods for deriving an EMT in field theory. The first is the \textit{Hilbert approach}\cite{feynman2018feynman}, which varies the action with respect to the spacetime metric $g_{\mu\nu}$, leading to what is often called the Hilbert or metric EMT. This method guarantees a symmetric, covariantly conserved tensor by construction, which is compatible with angular momentum conservation and serves as the source in general relativity. The second is the \textit{Noether approach}, based on the invariance of the action under infinitesimal spacetime translations. The resulting EMT is referred to as the canonical EMT.

However, canonical EMTs derived from Noether’s theorem are often neither symmetric nor gauge-invariant. For instance, the EMT derived from a naive application of Noether’s theorem to the electromagnetic field yields an asymmetric tensor that fails to respect gauge invariance. This is problematic since physical quantities like energy and momentum must be gauge-independent. Historically, this discrepancy is resolved by manually adding so-called ``improvement terms''~\cite{landau2013classical,freese2022noether}, but such modifications are mathematically artificial and lack a unified justification across field theories.

These challenges raise a fundamental question: \textit{Can the canonical EMT be constructed in a way that is naturally symmetric, gauge-invariant, and equivalent to the Hilbert EMT—without relying on ad hoc modifications?} More specifically, is there a unified and systematic framework that extends Noether’s theorem to yield a physically meaningful EMT across both gauge and gravitational theories?

This paper answers this question affirmatively by developing a geometric and variational approach that consistently relates Noether currents to the Hilbert energy–momentum tensor. We show that by properly treating the gauge potential as a differential 1-form rather than a mere vector field, one can derive a symmetric and gauge-invariant canonical EMT directly from Noether’s theorem. Moreover, by extending this formalism to general relativity using the vielbein and Palatini frameworks, we demonstrate that the canonical EMT derived from diffeomorphism invariance coincides with the Einstein tensor, and hence the Hilbert EMT.

The structure of the paper is as follows. In Sec.~\ref{sec:variational}, we introduce the variational framework for deriving canonical EMTs, emphasizing the role of both field variations and spacetime diffeomorphisms within Noether’s theorem. In Sec.~\ref{sec:GT}, we apply this framework to Abelian and non-Abelian gauge theories, demonstrating how a symmetric, gauge-invariant canonical EMT naturally arises. Unlike conventional treatments of gauge theory that presume a flat Minkowski background, our approach constructs the gauge-invariant energy–momentum tensor without assuming flat geometry. Section~\ref{sec:1-form} discusses the necessity of treating gauge potentials as 1-forms to ensure the consistency of the Noether procedure and to avoid the need for improvement terms. In Sec.~\ref{sec:GR}, we transition to general relativity by reviewing the Palatini variation, showing that metric compatibility implies a torsion-free connection, and recast the theory in the vielbein formalism to construct a coordinate-free Noether current. In Sec.~\ref{sec:Equivalence EMT}, we establish the equivalence between the canonical and Hilbert EMTs under symmetry-preserving conditions. Finally, in Sec.~\ref{sec:Einstein eq}, we derive Einstein’s field equations purely from Noether symmetry arguments, thus unifying the derivation of EMTs in gauge and gravitational contexts.

\section{The Variational Principle}
\renewcommand{\thesubsection}{\thesection.\Alph{subsection}}
\label{sec:variational}

We denote $x^\gamma$ as a point in spacetime and express the electromagnetic gauge potential as the differential 1-form $A_\nu(x^\gamma)\, dx^\nu$. We now examine the effect of variations on the gauge potential and on the spacetime coordinates. The total variation includes two contributions: the variation of the gauge potential,
\[
A_\nu \rightarrow \tilde{A}_\nu = A_\nu + \delta A_\nu,
\]
and the spacetime variation generated by a vector field $\delta x^\gamma$, given by
\[
x^\gamma \rightarrow \tilde{x}^\gamma = x^\gamma + \delta x^\gamma.
\]
The variation of the gauge 1-form then reads
\[
\tilde{A}_\nu(\tilde{x}^\gamma)\, d\tilde{x}^\nu - A_\nu(x^\gamma)\, dx^\nu = 
\left[ \delta A_\nu + 
\underbracket[0.4pt][0pt]{A_{\nu,\gamma} \delta x^\gamma + A_\gamma \delta x^\gamma{}_{,\nu}}_{\text{identified as } \hat{\mathcal{L}}_{\delta x} A} \right] dx^\nu + \mathcal{O}(\delta^2),
\]
where $\hat{\mathcal{L}}_{\delta x} A$ denotes the Lie derivative of the 1-form $A$ along the vector field $\delta x$. The total variation of the gauge potential is thus defined as
\begin{align*}
\Delta A_\nu &= \delta A_\nu + A_{\nu,\gamma} \delta x^\gamma + A_\gamma \delta x^\gamma{}_{,\nu}, \\
\Delta A &= \delta A + \hat{\mathcal{L}}_{\delta x} A.
\end{align*}

We now consider the Lagrangian density $\mathscr{L}[A_\nu, A_{\nu,\mu}, x^\gamma]$ and the associated action $S = \int d^4x\, \mathscr{L}$. The equation of motion (EoM) and Noether’s theorem follow from the variation of the action. The variation $\Delta S$ splits into two terms:
\[
\Delta S = \int \Delta d^4x \cdot \mathscr{L}_{EM} + \int d^4x \cdot \Delta \mathscr{L}_{EM},
\]
where the first term represents the variation of the volume form, given by \cite{ryder1996quantum}
\[
\Delta d^4x = \delta x^\gamma{}_{,\gamma} \cdot d^4x.
\]
The second term corresponds to the variation of the Lagrangian\cite{ryder1996quantum}:
\begin{align*}
\Delta \mathscr{L} &= \mathscr{L}[\tilde{A}_\nu(\tilde{x}^\gamma), \tilde{A}_{\nu,\mu}(\tilde{x}^\gamma), \tilde{x}^\gamma] 
- \mathscr{L}[A_\nu(x^\gamma), A_{\nu,\mu}(x^\gamma), x^\gamma] \\
&= \left[ \frac{\partial \mathscr{L}_{EM}}{\partial A_\nu} \, \delta A_\nu 
+ \frac{\partial \mathscr{L}}{\partial (\partial_\mu A_\nu)} \, \delta(\partial_\mu A_\nu) \right] 
+ \mathscr{L}_{,\gamma} \delta x^\gamma. \nonumber
\end{align*}

\noindent The full variation of the action then becomes:
\begin{adjustwidth}{-2cm}{-1.5cm}
\begin{align} \label{eq:Delta S_ab}
\Delta S &= \int \left[ \frac{\partial \mathscr{L}}{\partial A_\nu} 
- \partial_\mu \left( \frac{\partial \mathscr{L}}{\partial(\partial_\mu A_\nu)} \right) \right] \delta A_\nu \, d^4x + \int \left[ \partial_\mu \left( \frac{\partial \mathscr{L}}{\partial(\partial_\mu A_\nu)} \delta A_\nu \right) 
+ (\mathscr{L} \delta x^\gamma)_{,\gamma} \right] d^4x \nonumber \\
&= \int \left[ \frac{\partial \mathscr{L}}{\partial A_\nu} 
- \partial_\mu \left( \frac{\partial \mathscr{L}}{\partial(\partial_\mu A_\nu)} \right) \right] \delta A_\nu \, d^4x \nonumber 
+ \int \partial_\mu \left[ \frac{\partial \mathscr{L}}{\partial(\partial_\mu A_\nu)} \Delta A_\nu 
- \frac{\partial \mathscr{L}}{\partial(\partial_\mu A_\nu)} \left( A_{\nu,\gamma} \delta x^\gamma + A_\gamma \delta x^\gamma_{,\nu} \right) 
+ \delta^\mu_\gamma \mathscr{L} \delta x^\gamma \right] d^4x. \nonumber
\end{align}
\end{adjustwidth}

\noindent The equation of motion is given by:
\[
\partial_\mu \left( \frac{\partial \mathscr{L}}{\partial(\partial_\mu A_\nu)} \right) = \frac{\partial \mathscr{L}}{\partial A_\nu}.
\]

\noindent We also identify two associated conserved currents:
\begin{align*}
\partial_\mu \mathcal{J}^\mu_{\Delta A} &= \partial_\mu \left( \frac{\partial \mathscr{L}}{\partial(\partial_\mu A_\nu)} \, \Delta A_\nu \right), \\
\partial_\mu \mathcal{J}^\mu_{\delta x} &= \partial_\mu \left[ 
- \frac{\partial \mathscr{L}}{\partial(\partial_\mu A_\nu)} A_{\nu,\gamma} \delta x^\gamma 
- \frac{\partial \mathscr{L}}{\partial(\partial_\mu A_\nu)} A_\gamma \delta x^\gamma{}_{,\nu} 
+ \delta^\mu_\gamma \mathscr{L} \delta x^\gamma \right].
\end{align*}

\noindent With this preparation, we are ready to examine the Noether current associated with spacetime symmetry. The conserved current $\partial_\mu \mathcal{J}^\mu_{\delta x}$ will lead to the correct energy-momentum tensor for the electromagnetic field, Yang–Mills theory, and general relativity.

\section{Gauge Theory} 
\label{sec:GT}
\subsection{Electromagnetic Field (Abelian Gauge Theory)}

In this section, we do not assume a flat spacetime background; our construction remains valid on arbitrary curved manifolds. The Lagrangian density $\mathscr{L}_{EM}$ and the action $S$ of the electromagnetic field are given by
\begin{align*}
\mathscr{L}_{EM} = -\frac{1}{16\pi c} g^{\mu\alpha}g^{\nu\beta}F_{\alpha\beta}F_{\mu\nu}\sqrt{-g},
\end{align*}
where $g^{\mu\alpha}$ is the metric tensor, $g = \det(g_{\mu\nu})$ is the determinant of the metric, and $F_{\mu\nu} = A_{\nu,\mu} - A_{\mu,\nu}$ is the electromagnetic field strength. The Noether current associated with spacetime symmetry is given by
\begin{align*}
\partial_\mu \mathcal{J}^\mu_{\delta x-EM} 
= \partial_\mu \left[ 
- \frac{\partial \mathscr{L}_{EM}}{\partial(\partial_\mu A_\nu)} A_{\nu,\gamma} \delta x^\gamma 
- \underbracket[0.4pt][0pt]{\frac{\partial \mathscr{L}_{EM}}{\partial(\partial_\mu A_\nu)} A_\gamma \delta x^\gamma_{,\nu}}_{(*)} 
+ \delta^\mu_\gamma \mathscr{L}_{EM} \delta x^\gamma 
\right].
\end{align*}

\noindent Let us now evaluate the $(*)$ term:
\begin{align*}
\underbracket[0.4pt][0pt]{\partial_\mu \left[ \frac{\partial \mathscr{L}_{EM}}{\partial(\partial_\mu A_\nu)} A_{\gamma} \delta x^\gamma_{,\nu} \right]}_{(*)}
&= \underbracket[0.4pt][0pt]{\left[ \frac{\partial \mathscr{L}_{EM}}{\partial (\partial_\mu A_\nu)} A_\gamma \delta x^\gamma \right]_{,\nu \mu}}_{(*1)} 
- \partial_\mu \left[ \underbracket[0.4pt][0pt]{ \left( \frac{\partial \mathscr{L}_{EM}}{ \partial (\partial_\mu A_\nu)}\right)_{,\nu} A_\gamma \delta x^\gamma }_{(*2)} \right] 
- \partial_\mu \left[ \underbracket[0.4pt][0pt]{\frac{\partial \mathscr{L}_{EM}}{\partial(\partial_\mu A_\nu)} A_{\gamma,\nu} \delta x^\gamma}_{(*3)} \right].
\end{align*}

\noindent The $(\ast 1)$ term evaluates to
\begin{align} \label{eq:anti_ab}
\underbracket[0.4pt][0pt]{\left[ \frac{\partial \mathscr{L}_{EM}}{\partial (\partial_\mu A_\nu)} A_\gamma \delta x^\gamma \right]_{,\nu \mu}}_{(*1)} 
= \left[ \left( -\frac{1}{4\pi c} g^{\mu\alpha} g^{\nu\beta} F_{\alpha\beta} \sqrt{-g} \right) A_\gamma \delta x^\gamma \right]_{,\nu \mu} = 0,
\end{align}
due to the antisymmetry of $F_{\mu\nu}$ and the symmetry of partial derivatives with respect to $\nu$ and $\mu$.

\noindent The $(\ast 2)$ term becomes
\begin{align} \label{eq:*2_ab}
\underbracket[0.4pt][0pt]{ \left( \frac{\partial \mathscr{L}_{EM}}{ \partial (\partial_\mu A_\nu)}\right)_{,\nu} A_\gamma \delta x^\gamma }_{(*2)}
= \left( - \frac{\partial \mathscr{L}_{EM}}{ \partial (\partial_\nu A_\mu)} \right)_{,\nu} A_\gamma \delta x^\gamma
\underbracket{=}_{\text{EoM}} - \frac{\partial \mathscr{L}_{EM}}{\partial A_\mu} A_\gamma \delta x^\gamma = 0,
\end{align}
using the equations of motion.

\noindent Therefore, the $(*)$ term reduces to the third contribution only:
\[
\underbracket[0.4pt][0pt]{\partial_\mu\left[ \frac{\partial \mathscr{L}_{EM}}{\partial(\partial_\mu A_\nu)} A_\gamma \delta x^\gamma \right]_{,\nu}}_{(*)} 
= -\partial_\mu \left[ \underbracket[0.4pt][0pt]{ \frac{\partial \mathscr{L}_{EM}}{\partial(\partial_\mu A_\nu)} A_{\gamma,\nu} \delta x^\gamma }_{(*3)} \right].
\]

\noindent Substituting this back, the final expression for the divergence of the Noether current is
\begin{align*}
\partial_\mu \mathcal{J}^\mu_{\delta x-EM} 
&= \partial_\mu \left[ 
- \frac{\partial \mathscr{L}_{EM}}{\partial(\partial_\mu A_\nu)} A_{\nu,\gamma} \delta x^\gamma 
+ \underbracket[0.4pt][0pt]{ \frac{\partial \mathscr{L}_{EM}}{\partial(\partial_\mu A_\nu)} A_{\gamma,\nu} \delta x^\gamma }_{(*)=(3)} 
+ \delta^\mu_\gamma \mathscr{L}_{EM} \delta x^\gamma \right] \\
&= \partial_\mu \left[ \left( - \frac{\partial \mathscr{L}_{EM}}{\partial(\partial_\mu A_\nu)} \left( A_{\nu,\gamma} - A_{\gamma,\nu} \right) + \delta^\mu_\gamma \mathscr{L}_{EM} \right) \delta x^\gamma \right] \\
&= \partial_\mu \left[ \underbracket[0.4pt][0pt]{ \left( - \frac{\partial \mathscr{L}_{EM}}{\partial(\partial_\mu A_\nu)} F_{\gamma\nu} + \delta^\mu_\gamma \mathscr{L}_{EM} \right) }_{(t_{EM})^\mu_\gamma} \delta x^\gamma \right].
\end{align*}

\noindent Hence, we identify the energy-momentum tensor density $(t_{EM})^\mu_\gamma$ and the corresponding energy-momentum tensor $(T_{EM})^\mu_\gamma$ as
\begin{align*}
(t_{EM})^\mu_\gamma &= - \frac{\partial \mathscr{L}_{EM}}{\partial(\partial_\mu A_\nu)} F_{\gamma\nu} + \delta^\mu_\gamma \mathscr{L}_{EM} \\
&= \frac{1}{4\pi c} g^{\mu\alpha} g^{\nu\beta} F_{\alpha\beta} F_{\gamma\nu} \sqrt{-g} 
- \delta^\mu_\gamma \frac{1}{16\pi c} g^{\varepsilon\alpha} g^{\nu\beta} F_{\alpha\beta} F_{\varepsilon\nu} \sqrt{-g} \\
&= \underbracket[0.4pt][0pt]{ \left( \frac{1}{4\pi c} F^{\mu\nu} F_{\gamma\nu} - \delta^\mu_\gamma \frac{1}{16\pi c} F^{\alpha\beta} F_{\alpha\beta} \right) }_{(T_{EM})^\mu_\gamma} \sqrt{-g}.
\end{align*}

\noindent The resulting canonical EMT $(T_{EM})^\mu_\gamma$ is symmetric and gauge invariant. For clarity, we summarize the relation as $(t_{EM})^\mu_\gamma = (T_{EM})^\mu_\gamma \sqrt{-g}$.

\subsection{Yang--Mills Theory (Non-Abelian Gauge Theory)}

In Yang--Mills theory, the gauge potential becomes a Lie algebra-valued 1-form,
\[
\mathbf{A}_\nu \, dx^\nu = A^a_\nu \hat{T}_a \, dx^\nu,
\]
where $\hat{T}_a$ denotes the generators of the Lie algebra. The field strength in local coordinates is given by
\begin{align*} \label{eq:Guv}
\mathbf{F}_{\mu\nu} 
&= \partial_\mu \mathbf{A}_\nu - \partial_\nu \mathbf{A}_\mu + [\mathbf{A}_\mu, \mathbf{A}_\nu] \\
&= \hat{T}_a 
\underbracket[0.4pt][0pt]{\left( \partial_\mu A_\nu^a - \partial_\nu A_\mu^a + f_{bc}^a A_\mu^b A_\nu^c \right)}_{ \equiv F_{\mu\nu}^a},
\end{align*}
where the Lie algebra structure constants satisfy
\[
[\hat{T}_a, \hat{T}_b] = f_{ab}^c \hat{T}_c.
\]
The Killing form is denoted as
\[
K_{ab} = \mathrm{Tr}(\hat{T}_a \hat{T}_b).
\]
The Yang--Mills Lagrangian is then
\begin{align} \label{eq:Lagrangian}
\mathscr{L}_{YM} 
&= \mathrm{Tr} \left( -\frac{1}{16\pi c} g^{\mu\alpha} g^{\nu\beta} \mathbf{F}_{\mu\nu} \mathbf{F}_{\alpha\beta} \sqrt{-g} \right) 
= -\frac{1}{16\pi c} K_{ab} g^{\mu\alpha} g^{\nu\beta} F^a_{\mu\nu} F^b_{\alpha\beta} \sqrt{-g}.
\end{align}
The equation of motion (EoM) takes the standard form
\[
\partial_\mu \left( \frac{\partial \mathscr{L}_{YM}}{\partial(\partial_\mu A^a_\nu)} \right) = \frac{\partial \mathscr{L}_{YM}}{\partial A^a_\nu}.
\]
The Noether current associated with spacetime symmetry is
\begin{align} \label{eq:YM_J}
\partial_\mu \mathcal{J}^\mu_{\delta x-YM} 
= \partial_\mu \left[
- \frac{\partial \mathscr{L}_{YM}}{\partial(\partial_\mu A^a_\nu)} A^a_{\nu,\gamma} \delta x^\gamma
- \underbracket[0.4pt][0pt]{\frac{\partial \mathscr{L}_{YM}}{\partial(\partial_\mu A^a_\nu)} A^a_\gamma \delta x^\gamma_{,\nu}}_{(*)}
+ \delta^\mu_\gamma \mathscr{L}_{YM} \delta x^\gamma
\right].
\end{align}

\noindent Evaluating the $(*)$ term:
\begin{align*} 
\underbracket[0.4pt][0pt]{\partial_\mu \left[ \frac{\partial \mathscr{L}_{YM}}{\partial(\partial_\mu A^a_\nu)} A^a_{\gamma} \delta x^\gamma_{,\nu} \right]}_{(*)}
= \underbracket[0.4pt][0pt]{\left[ \frac{\partial \mathscr{L}_{YM}}{\partial (\partial_\mu A^a_\nu)} A^a_\gamma \delta x^\gamma \right]_{,\nu \mu}}_{(*1)} 
- \partial_\mu \left[ \underbracket[0.4pt][0pt]{\left( \frac{\partial \mathscr{L}_{YM}}{ \partial (\partial_\mu A^a_\nu)} \right)_{,\nu} A^a_\gamma \delta x^\gamma}_{(*2)} \right] 
- \partial_\mu \left[ \underbracket[0.4pt][0pt]{\frac{\partial \mathscr{L}_{YM}}{\partial(\partial_\mu A^a_\nu)} A^a_{\gamma,\nu} \delta x^\gamma}_{(*3)} \right].
\end{align*}

\noindent The first term vanishes due to antisymmetry:
\begin{align} \label{eq:1_YM}
\left( \frac{\partial \mathscr{L}_{YM}}{\partial (\partial_\mu A_\nu^a)} A_\gamma^a \delta x^\gamma \right)_{,\nu\mu}
= \left[ \left( -\frac{1}{4\pi c} K_{ab} g^{\alpha\mu} g^{\beta\nu} F_{\alpha\beta}^b \sqrt{-g} \right) A_\gamma^a \delta x^\gamma \right]_{,\nu\mu} = 0
\end{align}
analogous to Eq.~\eqref{eq:anti_ab} in the Abelian case. The $(\ast 2)$ term involves structure constants:
\begin{align*} \label{eq:(*2)}
\left( \frac{\partial \mathscr{L}_{YM}}{ \partial (\partial_\mu A^a_\nu)} \right)_{,\nu} A^a_\gamma \delta x^\gamma
= \left( - \frac{\partial \mathscr{L}_{YM}}{ \partial (\partial_\nu A^a_\mu)} \right)_{,\nu} A^a_\gamma \delta x^\gamma
\underbracket{=}_{\text{EoM}}
- \frac{\partial \mathscr{L}_{YM}}{ \partial A^a_\mu } A^a_\gamma \delta x^\gamma
= -\frac{\partial \mathscr{L}_{YM}}{\partial(\partial_\mu A^a_\nu)} f_{bn}^a A^b_\gamma A^n_\nu \delta x^\gamma.
\end{align*}

\noindent Therefore, the $(*)$ term becomes:
\[
\underbracket[0.4pt][0pt]{\partial_\mu \left[ \frac{\partial \mathscr{L}_{YM}}{\partial(\partial_\mu A^a_\nu)} A^a_{\gamma} \delta x^\gamma_{,\nu} \right]}_{(*)}
= -\partial_\mu \left[ \underbracket[0.4pt][0pt]{ -\frac{\partial \mathscr{L}_{YM}}{\partial(\partial_\mu A^a_\nu)} f_{bn}^a A^b_\gamma A^n_\nu \delta x^\gamma }_{(*2)} \right]
- \partial_\mu \left[ \underbracket[0.4pt][0pt]{ \frac{\partial \mathscr{L}_{YM}}{\partial(\partial_\mu A^a_\nu)} A^a_{\gamma,\nu} \delta x^\gamma }_{(*3)} \right].
\]

\noindent Substituting back into Eq. \eqref{eq:YM_J}, the conserved Noether current becomes:
\begin{align*}
\partial_\mu \mathcal{J}^\mu_{\delta x-YM} 
&= \partial_\mu \left[ 
- \frac{\partial \mathscr{L}_{YM}}{\partial(\partial_\mu A^a_\nu)} A^a_{\nu,\gamma} \delta x^\gamma 
+ \underbracket[0.4pt][0pt]{ \frac{\partial \mathscr{L}_{YM}}{\partial(\partial_\mu A^a_\nu)} A^a_{\gamma,\nu} \delta x^\gamma }_{(*3)}
- \underbracket[0.4pt][0pt]{ \frac{\partial \mathscr{L}_{YM}}{\partial(\partial_\mu A^a_\nu)} f_{bn}^a A^b_\gamma A^n_\nu \delta x^\gamma }_{(*2)}
+ \delta^\mu_\gamma \mathscr{L}_{YM} \delta x^\gamma
\right] \\
&= \partial_\mu \left[ 
\left( - \frac{\partial \mathscr{L}_{YM}}{\partial(\partial_\mu A^a_\nu)}
\left( A^a_{\nu,\gamma} - A^a_{\gamma,\nu} + f_{bn}^a A^b_\gamma A^n_\nu \right)
+ \delta^\mu_\gamma \mathscr{L}_{YM} \right)
\delta x^\gamma
\right] \\
&= \partial_\mu \left[
\underbracket[0.4pt][0pt]{\left( - \frac{\partial \mathscr{L}_{YM}}{\partial(\partial_\mu A^a_\nu)} F^a_{\gamma\nu} + \delta^\mu_\gamma \mathscr{L}_{YM} \right)}_{(t_{YM})^\mu_\gamma}
\delta x^\gamma
\right].
\end{align*}

\noindent We thus identify the energy-momentum tensor density $(t_{YM})^\mu_\gamma$ and the corresponding energy-momentum tensor $(T_{YM})^\mu_\gamma$ as
\begin{align*}
(t_{YM})^\mu_\gamma 
&= - \frac{\partial \mathscr{L}_{YM}}{\partial(\partial_\mu A^a_\nu)} F^a_{\gamma\nu} + \delta^\mu_\gamma \mathscr{L}_{YM} \\
&= \left( \frac{1}{4\pi c} K_{ab} g^{\alpha\mu} g^{\beta\nu} F^b_{\alpha\beta} \sqrt{-g} \right) F^a_{\gamma\nu} 
+ \delta^\mu_\gamma \left( -\frac{1}{16\pi c} K_{ab} g^{\mu\alpha} g^{\nu\beta} F^a_{\mu\nu} F^b_{\alpha\beta} \sqrt{-g} \right) \\
&= \underbracket[0.4pt][0pt]{\left( \frac{1}{4\pi c} F_a^{\mu\nu} F^a_{\gamma\nu} - \delta^\mu_\gamma \frac{1}{16\pi c} F^{\alpha\beta}_a F^a_{\alpha\beta} \right)}_{(T_{YM})^\mu_\gamma} \sqrt{-g}.
\end{align*}
. The result is the same as in the Abelian case.

\section{The 1-form Nature of the Gauge Potential}
\label{sec:1-form}

\noindent From a rigorous mathematical perspective, particularly within the differential geometric framework used to formulate gauge theories, the fundamental object representing the gauge field is a \textbf{connection 1-form} \(\mathbf{A}\). This connection lives on a principal bundle (or an associated vector bundle) over spacetime and encodes how internal gauge degrees of freedom are related at infinitesimally separated spacetime points. It defines the notion of parallel transport for fields charged under the gauge group.

In physics, gauge potentials are often expressed either as vector fields \(\mathbf{A}^\mu \partial_\mu\) or as covariant 1-forms \(\mathbf{A}_\mu dx^\mu\), which are related through the spacetime metric:
\begin{align*}
\mathbf{A}^\mu &= g^{\mu\nu} \mathbf{A}_\nu,
\end{align*}
with the corresponding contravariant field strength tensor given by
\begin{align*}
\mathbf{F}^{\mu\nu} &= g^{\mu\alpha} g^{\nu\beta} \mathbf{F}_{\alpha\beta}.
\end{align*}

\noindent The electromagnetic (or Yang–Mills) Lagrangian can be expressed in two equivalent forms:
\begin{align*}
\mathscr{L}[\mathbf{A}_\mu] &= -\frac{1}{16\pi c} g^{\mu\alpha} g^{\nu\beta} \mathbf{F}_{\alpha\beta} \mathbf{F}_{\mu\nu}, \\
\mathscr{L}[\mathbf{A}^\mu] &= -\frac{1}{16\pi c} g_{\mu\alpha} g_{\nu\beta} \mathbf{F}^{\alpha\beta} \mathbf{F}^{\mu\nu}.
\end{align*}
These forms yield the same equations of motion under variation with respect to \(\mathbf{A}_\mu\) or \(\mathbf{A}^\mu\). However, when applying Noether’s theorem to derive the canonical energy-momentum tensor, only the 1-form formulation based on \(\mathbf{A}_\mu\) produces the correct result.

The subtle distinction lies in how each formulation transforms under spacetime symmetries, where the Lie derivative governs the total variation:
\begin{align*}
\Delta \mathbf{A}_\nu &= \delta \mathbf{A}_\nu + \underbracket[0.4pt][0pt]{\mathbf{A}_{\nu,\gamma} \delta x^\gamma + \mathbf{A}_\gamma \, {\delta x^\gamma}_{,\nu}}_{\text{Lie deriv. of a 1-form}}, \\
\Delta \mathbf{A}^\nu &= \delta \mathbf{A}^\nu + \underbracket[0.4pt][0pt]{{\mathbf{A}^\nu}_{,\gamma} \delta x^\gamma - \mathbf{A}^\gamma \, {\delta x^\nu}_{,\gamma}}_{\text{Lie deri. of a vector field}}.
\end{align*}

\noindent Following the Noether procedure, the final term in \(\Delta \mathbf{A}^\nu\) obstructs the correct derivation of the canonical energy-momentum tensor. This reinforces both the geometric and physical significance of choosing the covariant 1-form \(\mathbf{A}_\mu\) as the fundamental gauge field variable. Noether’s theorem thus naturally reveals the 1-form structure of the gauge potential.

\section{General Relativity}
\label{sec:GR}
\subsection{The Stress-Energy-Momentum Pseudotensor}

The stress-energy-momentum pseudotensor is a traditional method to describe the concept of energy-momentum of the gravitational field. The Landau–Lifshitz pseudotensor \( t^{\mu\nu}_{LL} \) is derived from the Einstein field equations \( \mathbb{G}^{\mu\nu} = \kappa T^{\mu\nu} \) to satisfy the conservation law \cite{landau2013classical}
\[
(T^{\mu\nu} + t^{\mu\nu}_{LL})_{,\nu} = 0,
\]
where \( \mathbb{G}^{\mu\nu} \) is the Einstein tensor, \( \kappa = \frac{8\pi G}{c^4} \), and \( T^{\mu\nu} \) is the energy-momentum tensor of the matter source.

\begin{align*}
t^{\mu\nu}_{LL} = -\frac{1}{2\kappa} \mathbb{G}^{\mu\nu} + \frac{1}{2\kappa(-g)} \left[ (-g)(g^{\mu\nu}g^{\alpha\beta} - g^{\mu\alpha}g^{\nu\beta}) \right]_{,\alpha\beta}
\end{align*}

\noindent The pseudotensor \( t^{\mu\nu}_{LL} \) is symmetric, but depends explicitly on the Christoffel symbols—i.e., it is coordinate dependent and can vanish in specific coordinate systems.

\medskip

The Dirac pseudotensor \( t^{\mu\nu}_D \) is derived using the standard Noether's theorem starting from an equivalent action \( \mathscr{L}_{GR}^* \) \cite{dirac1996general}. The original Einstein–Hilbert action is given by
\[
\mathscr{L}_{GR} = R\sqrt{-g} = \mathscr{L}_{GR}[g^{\mu\nu}, \partial_\gamma g^{\mu\nu}, \partial_\gamma\partial_\eta g^{\mu\nu}],
\]
which depends on the second derivative of the metric tensor. Here, \( R \) is the Ricci scalar and \( g = \det(g_{\mu\nu}) \). The Ostrogradsky instability suggests that a Lagrangian depending on higher-order derivatives (beyond first order) leads to instability\cite{Ostrogradsky:1850fid} \cite{Woodard:2015zca}. To avoid this, one can use an equivalent action that depends only on first derivatives:
\begin{align*}
\mathscr{L}_{GR}^* &= \mathscr{L}_{GR} - \partial_\mu \left( \sqrt{-g}g^{\mu\nu} \Gamma^\sigma_{\nu\sigma} - \sqrt{-g}g^{\sigma\nu} \Gamma^\mu_{\nu\sigma} \right) \\
&= \sqrt{-g}g^{\mu\nu} \left( \Gamma^\tau_{\mu\nu} \Gamma^\sigma_{\tau\sigma} - \Gamma^\tau_{\mu\sigma} \Gamma^\sigma_{\tau\nu} \right) \\
&= \mathscr{L}_{GR}^*[g^{\mu\nu}, \partial_\gamma g^{\mu\nu}]
\end{align*}
. This action yields the same equations of motion but only depends on first derivatives of the metric, allowing the application of standard Noether analysis. However, \( \mathscr{L}_{GR}^* \) is no longer a scalar under general coordinate transformations. The resulting Dirac pseudotensor is:
\begin{align*}
t^{\mu\nu}_D = \frac{1}{2\kappa(-g)} \left[
g^{\mu\gamma} \left( g^{\alpha\beta} \sqrt{-g} \right)_{,\gamma}
\left( \Gamma^\nu_{\alpha\beta} - \delta^\nu_\beta \Gamma^\sigma_{\alpha\sigma} \right)
- g^{\mu\nu} g^{\alpha\beta}
\left( \Gamma^\rho_{\alpha\beta} \Gamma^\sigma_{\rho\sigma}
- \Gamma^\rho_{\alpha\sigma} \Gamma^\sigma_{\beta\rho} \right)
\right]
\end{align*}

\noindent Like \( t^{\mu\nu}_{LL} \), the Dirac pseudotensor is coordinate dependent and can vanish in specific coordinate systems. Additionally, it lacks symmetry in its indices.

\medskip

As we have previously derived the symmetric and gauge-invariant canonical energy-momentum tensor for abelian and non-abelian gauge fields, we now extend the construction to general relativity. To preserve the scalar property of the Lagrangian while avoiding higher-order derivatives, we adopt the Palatini variation for deriving the field equations. Furthermore, to apply Noether's theorem appropriately, we employ the vielbein formalism in order to formulate the conservation law.

\subsection{Review of the Palatini Variation}

The Palatini variation\cite{palatini1919deduzione} treats the metric tensor \( g^{\omega\sigma} \) and the affine connection \( \Gamma^\varepsilon_{\kappa\gamma} \) as independent fields. The Riemann curvature tensor is defined as
\begin{align} \label{eq:R_curvature}
R^\varepsilon_{\kappa\omega\sigma} = \Gamma^\varepsilon_{\kappa\sigma,\omega} - \Gamma^\varepsilon_{\kappa\omega,\sigma} + \Gamma^\varepsilon_{\gamma\omega}\Gamma^\gamma_{\kappa\sigma} - \Gamma^\varepsilon_{\gamma\sigma}\Gamma^\gamma_{\kappa\omega},
\end{align}
and the torsion tensor is given by
\begin{align} \label{eq:torsion}
T^\alpha_{\beta\gamma} = \Gamma^\alpha_{\beta\gamma} - \Gamma^\alpha_{\gamma\beta} = - T^\alpha_{\gamma\beta}.
\end{align}

\noindent In the following, we do \emph{not} assume the connection to be torsion-free. That is, in general \( T^\alpha_{\beta\gamma} \neq 0 \), which implies \( \Gamma^\alpha_{\beta\gamma} \neq \Gamma^\alpha_{\gamma\beta} \).

The Einstein–Hilbert Lagrangian takes the form
\begin{align*}
\mathscr{L}_{GR} = \frac{1}{2\kappa} \sqrt{-g} g^{\mu\nu} \delta^\omega_\varepsilon R^\varepsilon_{\kappa\omega\sigma},
\end{align*}
and the corresponding action is
\begin{align*}
S = \frac{1}{2\kappa} \int d^4x\, \sqrt{-g} g^{\kappa\sigma} \delta^\omega_\varepsilon R^\varepsilon_{\kappa\omega\sigma}.
\end{align*}

\noindent We recall that the Ricci tensor and scalar curvature are defined as
\begin{align*}
R_{\kappa\sigma} &= \delta^\omega_\varepsilon R^\varepsilon_{\kappa\omega\sigma}, \\
R &= g^{\kappa\sigma} R_{\kappa\sigma},
\end{align*}
so that the Einstein–Hilbert Lagrangian can be compactly written as \( \mathscr{L}_{GR} = \frac{1}{2\kappa} \sqrt{-g} R \).

Since \( g^{\mu\nu} \) and \( \Gamma^\alpha_{\mu\nu} \) are treated as independent fields, the Lagrangian depends only on first derivatives of the connection. The Palatini variation yields
\begin{align} \label{eq:Action_g}
\delta S &= \int \left( \frac{\partial \mathscr{L}_{GR}}{\partial g^{\mu\nu}} \delta g^{\mu\nu} + \frac{\partial \mathscr{L}_{GR}}{\partial \Gamma^\alpha_{\mu\nu}} \delta \Gamma^\alpha_{\mu\nu} + \frac{\partial \mathscr{L}_{GR}}{\partial \Gamma^\alpha_{\mu\nu,\gamma}} \delta \Gamma^\alpha_{\mu\nu,\gamma} \right) d^4x \nonumber \\
&= \int \underbracket[0.4pt][0pt]{\frac{\partial \mathscr{L}_{GR}}{\partial g^{\mu\nu}}}_{\text{EoM \#1}} \delta g^{\mu\nu} \, d^4x + \int \underbracket[0.4pt][0pt]{\left[ \frac{\partial \mathscr{L}_{GR}}{\partial \Gamma^\alpha_{\mu\nu}} - \left( \frac{\partial \mathscr{L}_{GR}}{\partial \Gamma^\alpha_{\mu\nu,\gamma}} \right)_{,\gamma} \right]}_{\text{EoM \#2}} \delta \Gamma^\alpha_{\mu\nu} \, d^4x + \int \left[ \frac{\partial \mathscr{L}_{GR}}{\partial \Gamma^\alpha_{\mu\nu,\gamma}} \delta \Gamma^\alpha_{\mu\nu} \right]_{,\gamma} d^4x.
\end{align}

\noindent This yields two equations of motion:
\begin{align*}
\frac{\partial \mathscr{L}_{GR}}{\partial g^{\mu\nu}} &= 0, \\
\frac{\partial \mathscr{L}_{GR}}{\partial \Gamma^\alpha_{\mu\nu}} &= \left( \frac{\partial \mathscr{L}_{GR}}{\partial \Gamma^\alpha_{\mu\nu,\gamma}} \right)_{,\gamma}.
\end{align*}

\subsubsection{EoM\#1 — Einstein Field Equation}

\noindent Since the curvature tensor does not depend on the metric, the first equation of motion becomes
\begin{align} \label{eq:Einstein}
\frac{\partial \mathscr{L}_{GR}}{\partial g^{\mu\nu}} = \frac{1}{2\kappa} \sqrt{-g} \left( R_{\mu\nu} - \frac{1}{2} g_{\mu\nu} R \right) = 0.
\end{align}

\noindent Define the Einstein tensor as
\[
\mathbb{G}_{\mu\nu} \equiv R_{\mu\nu} - \frac{1}{2}g_{\mu\nu}R.
\]
Then Eq.~\eqref{eq:Einstein} implies that, in the absence of matter, the Einstein field equations reduce to
\[
\mathbb{G}_{\mu\nu} = 0.
\]

\noindent When matter is present, the total Lagrangian becomes
\[
\mathscr{L}_{\text{total}} = \mathscr{L}_{GR} + \mathscr{L}_M,
\]
where \( \mathscr{L}_M \) denotes the matter contribution. Variation with respect to the metric gives
\begin{align*}
\frac{\partial \mathscr{L}_{\text{total}}}{\partial g^{\mu\nu}} 
= \frac{\partial \mathscr{L}_{GR}}{\partial g^{\mu\nu}} + \frac{\partial \mathscr{L}_M}{\partial g^{\mu\nu}} 
= \frac{1}{2\kappa} \sqrt{-g} \, \mathbb{G}_{\mu\nu} + \frac{\partial \mathscr{L}_M}{\partial g^{\mu\nu}} = 0.
\end{align*}
This implies
\[
\mathbb{G}_{\mu\nu} = \kappa \left( -\frac{2}{\sqrt{-g}} \frac{\partial \mathscr{L}_M}{\partial g^{\mu\nu}} \right) \equiv \kappa T^{(M)}_{\mu\nu},
\]
where \( T^{(M)}_{\mu\nu} \) is defined as the \textbf{Hilbert energy-momentum tensor} of the matter.

\subsubsection{EoM\#2 — Metric Compatibility and Torsion}

The second equation of motion leads to a relationship between the connection and the metric:
\begin{align} \label{eq:EoM2_result}
\left( \frac{\partial \mathscr{L}_{GR}}{\partial \Gamma^\alpha_{\mu\nu,\gamma}} \right)_{,\gamma} &= \frac{\partial \mathscr{L}_{GR}}{\partial \Gamma^\alpha_{\mu\nu}} \nonumber \\
\Rightarrow \quad
{(\sqrt{-g} \, g^{\mu\nu})_{,\alpha}} - (\sqrt{-g} \, g^{\mu\gamma})_{,\gamma} \delta^\nu_\alpha &= \sqrt{-g} \left(
g^{\kappa\sigma} \delta^\nu_\alpha \Gamma^\mu_{\kappa\sigma}
+ g^{\mu\nu} \Gamma^\omega_{\alpha\omega}
- g^{\kappa\nu} \Gamma^\mu_{\kappa\alpha}
- g^{\mu\sigma} \Gamma^\nu_{\alpha\sigma}
\right).
\end{align}

\noindent Solving Eq.~\eqref{eq:EoM2_result} gives the covariant derivative of the metric:
\[
\underbracket[0.4pt][0pt]{{g^{\mu\nu}}_{,\alpha}+g^{\eta\nu}\Gamma^\mu_{\eta\gamma}+g^{\mu\eta}\Gamma^\nu_{\eta\gamma}}_{\equiv{g^{\mu\nu}}_{;\alpha}}
=
\frac{1}{3} g^{\mu\gamma} T^\eta_{\eta\gamma} \, \delta^\nu_\alpha 
+ \frac{1}{3} g^{\mu\gamma} g^{\mu\nu} T^\gamma_{\alpha\gamma}
+ g^{\mu\kappa} T^\nu_{\kappa\alpha}.
\]

\noindent This relation shows the mutual implication between torsion and metric compatibility. We summarize it as follows:
\noindent This relation shows the mutual implication between torsion and metric compatibility. We summarize it as follows:

\begin{center}
\begin{tabular}{|p{0.45\textwidth}|p{0.45\textwidth}|}
\hline
Torsion-free then metric compatible & Metric compatible then torsion-free \\
\hline
\begin{minipage}[t]{0.45\textwidth}
    \centering
    If torsion-free:\, $T^\alpha_{\beta\gamma}=0$, then  ${g^{\mu\nu}}_{;\alpha}=0$ 
\end{minipage}
&
\begin{minipage}[t]{0.45\textwidth}
    \centering
    If Metric compatible:\, ${g^{\mu\nu}}_{;\alpha}=0$, then
    $T^\alpha_{\beta\gamma} = 0$
\end{minipage} \\
\hline
\end{tabular}
\end{center}

\noindent We conclude that, under the Palatini variation, the connection is metric-compatible \emph{if and only if} it is torsion-free.

\subsection{Noether's Variation Yields the Einstein Tensor}
\subsubsection{Review on Vielbein Formalism}

\noindent In the previous section, we derived the variation of the Lie algebra-valued gauge 1-form (gauge connection) $\mathbf{A}_\nu$ on a principal bundle in local coordinates as:
\begin{align} \label{eq:varaition_gauge}
\Delta A^a_{\nu} = \delta A^a_{\nu} + \underbrace{ A^a_{\nu,\gamma} \delta x^\gamma + A^a_{\gamma} \delta {x^\gamma}_{,\nu} }_{\hat{\mathcal{L}}_{\delta x} \mathbf{A}}
\end{align}

\noindent However, the situation is subtler when considering connections on the tangent bundle (see footnote 9 of \cite{yang1996symmetry}). Unlike gauge connections, which take values in internal symmetry groups, the affine connection $\Gamma^\alpha_{\mu\nu}$ takes values in the tangent space itself. Directly applying the Lie derivative to this tangent-space-valued object gives\cite{yano2020theory}:
\begin{align*}
\left(\hat{\mathcal{L}}_{\delta x} \Gamma\right)^\alpha_{\mu\nu} = {\delta x}^\varepsilon \Gamma^\alpha_{\mu\nu,\varepsilon} + \Gamma^\alpha_{\mu\varepsilon}{\delta x}^\varepsilon_{,\nu} + \underbracket[0.4pt][0pt]{{{\delta x}^\alpha_{,\mu\nu}} - \Gamma^\varepsilon_{\mu\nu}\delta x^\alpha_{,\varepsilon} + \Gamma^\alpha_{\varepsilon\nu}{\delta x}^\varepsilon_{,\mu}}_{\text{Due to tangent-space index structure}}
\end{align*}

\noindent To handle this properly, we adopt the vielbein formalism\cite{eguchi1980gravitation}. In this framework, the affine connection $\Gamma^\varepsilon_{\mu\nu}$ is replaced by the spin connection $\omega^a_{b\sigma}$, a $\mathfrak{gl}(4)$-valued 1-form on the frame bundle. The local basis $\{ \partial_\mu \}$ is replaced by orthonormal frames $\{ \hat{e}_a \}$ with the transformation:
\begin{align*}
\partial_\mu &= e^a_\mu \hat{e}_a \\
g_{\mu\nu} &= e^a_\mu e^b_\nu \eta_{ab} \\
e^\mu_a e^a_\nu &= \delta^\mu_\nu, \quad e^\mu_a e^b_\mu = \delta^b_a
\end{align*}

\noindent The spin connection $\omega^a_{b\sigma}$ is related to the affine connection by:
\begin{align*}
\omega^a_{b\sigma} = e^a_\varepsilon e^\kappa_b \Gamma^\varepsilon_{\kappa\sigma} + e^a_\mu \partial_\sigma e^\mu_b
\end{align*}

\noindent The curvature of the frame bundle, denoted $\mathscr{R}^a_{b\omega\sigma}$, is given by:
\begin{align*}
\mathscr{R}^a_{b\omega\sigma} = \omega^a_{b\sigma,\omega} - \omega^a_{b\omega,\sigma} + \omega^a_{c\omega} \omega^c_{b\sigma} - \omega^a_{c\sigma} \omega^c_{b\omega} = e^a_\varepsilon e^\kappa_b R^\varepsilon_{\kappa\omega\sigma}
\end{align*}

\noindent With these preparations, the Einstein-Hilbert action takes the form:
\begin{align} \label{eq:action_G}
S = \frac{1}{2\kappa} \int d^4x \sqrt{-g} \, g^{\kappa\sigma} \delta^\omega_\varepsilon R^\varepsilon_{\kappa\omega\sigma}
= \frac{1}{2\kappa} \int d^4x \, e \, \eta^{ae} e^\sigma_e e^\omega_c \mathscr{R}^c_{a\omega\sigma}
\end{align}
where \( e = \det(e^a_\mu) = \sqrt{-g} \) is the determinant of the vielbein.

\noindent The variation of this action follows a pattern analogous to Eq.~\eqref{eq:Action_g}:
\begin{align} \label{eq:Action_e}
\delta S &= \int \left( \frac{\partial \mathscr{L}_{GR}}{\partial e^a_\mu} \delta e^a_\mu 
+ \frac{\partial \mathscr{L}_{GR}}{\partial \omega^b_{c\nu}} \delta \omega^b_{c\nu} 
+ \frac{\partial \mathscr{L}_{GR}}{\partial (\omega^b_{c\nu,\gamma})} \delta \omega^b_{c\nu,\gamma} \right) d^4x \nonumber \\
&= \int \underbracket[0.4pt][0pt]{\frac{\partial \mathscr{L}_{GR}}{\partial e^a_\mu}}_{\text{EoM \#3}} \delta e^a_\mu \, d^4x 
+ \int \underbracket[0.4pt][0pt]{\left( \frac{\partial \mathscr{L}_{GR}}{\partial \omega^b_{c\nu}} - \left( \frac{\partial \mathscr{L}_{GR}}{\partial \omega^b_{c\nu,\gamma}} \right)_{,\gamma} \right)}_{\text{EoM \#4}} \delta \omega^b_{c\nu} \, d^4x 
+ \int \left[ \frac{\partial \mathscr{L}_{GR}}{\partial (\omega^b_{c\nu,\gamma})} \delta \omega^b_{c\nu} \right]_{,\gamma} d^4x
\end{align}

\noindent Since the metric is given by \( g_{\mu\nu} = \eta_{ab} e^a_\mu e^b_\nu \), its variation yields \( \delta g_{\mu\nu} = 2 \eta_{ab} e^a_\mu \delta e^b_\nu \). Therefore, the equation of motion from varying the vielbein (EoM~\#3) reproduces the Einstein field equation (EoM~\#1), while the equation of motion from varying the spin connection (EoM~\#4) corresponds to the condition for metric compatibility, analogous to EoM~\#2.

\subsubsection{Einstein Tensor Derived from Noether's Theorem}

\noindent We now apply Noether’s variation principle to the gravitational action:
\begin{align*} 
    \Delta S = \int \left[ \text{EoM terms} \right] d^4 x 
    + \int \left[ 
        \partial_\mu\left(
            \frac{\partial \mathscr{L}_{GR}}{\partial \omega^b_{c\nu,\mu}} \delta\omega^b_{c\nu}
        \right) 
        + (\mathscr{L}_{GR} \delta x^\gamma)_{,\gamma}
    \right] d^4 x 
\end{align*}

\noindent The spin connection \( \omega^b_{c\nu} \), being a $\mathfrak{gl}$-valued 1-form, transforms under infinitesimal diffeomorphisms according to the general Lie derivative rule, as in the gauge theory case [cf.~Eq.~\eqref{eq:varaition_gauge}]:
\begin{align} 
\Delta \omega^b_{c\nu} 
= \delta \omega^b_{c\nu} + 
\underbrace{
    \delta x^\varepsilon \partial_\varepsilon \omega^b_{c\nu} 
    + \omega^b_{c\varepsilon} \partial_\nu \delta x^\varepsilon
}_{\hat{\mathcal{L}}_{\delta x} \boldsymbol{\omega}} 
\end{align}

\noindent Consequently, the Noether current associated with spacetime translation is given by:
\begin{align*} 
\partial_\mu \mathcal{J}^\mu_{\delta x\text{-GR}} 
= \partial_\mu \left[ 
    -\frac{\partial \mathscr{L}_{GR}}{\partial \omega^b_{c\nu,\mu}} \omega^b_{c\nu,\gamma} \delta x^\gamma  
    - \underbracket[0.4pt][0pt]{\frac{\partial \mathscr{L}_{GR}}{\partial \omega^b_{c\nu,\mu}} \omega^b_{c\gamma} \delta {x^\gamma}_{,\nu}}_{(*)}
    + \delta^\mu_\gamma \mathscr{L}_{GR} \delta x^\gamma 
\right]
\end{align*}

\noindent We now compute the term marked by \((*)\):
\begin{align*}
\underbracket[0.4pt][0pt]{ \partial_\mu \left[ 
    \frac{\partial \mathscr{L}_{GR}}{\partial \omega^b_{c\nu,\mu}} \omega^b_{c\gamma} \delta {x^\gamma}_{,\nu}
\right]}_{(*)}
= 
\underbracket[0.4pt][0pt]{\left[
    \frac{\partial \mathscr{L}_{GR}}{\partial \omega^b_{c\nu,\mu}} \omega^b_{c\gamma} \delta x^\gamma 
\right]_{,\mu\nu}}_{(*1)}
- \partial_\mu \left[
    \underbracket[0.4pt][0pt]{\left(
        \frac{\partial \mathscr{L}_{GR}}{\partial \omega^b_{c\nu,\mu}}
    \right)_{,\nu} \omega^b_{c\gamma} \delta x^\gamma}_{(*2)}
\right]
- \partial_\mu \left[
    \underbracket[0.4pt][0pt]{\frac{\partial \mathscr{L}_{GR}}{\partial \omega^b_{c\nu,\mu}} \omega^b_{c\gamma,\nu} \delta x^\gamma}_{(*3)}
\right]
\end{align*}

\noindent The term \((*1)\) vanishes by analogy with Eq.~\eqref{eq:anti_ab} in the Abelian case and Eq.~\eqref{eq:1_YM} in the non-Abelian case:
\begin{align*}
\underbracket[0.4pt][0pt]{\left[
    \frac{\partial \mathscr{L}_{GR}}{\partial \omega^b_{c\nu,\mu}} \omega^b_{c\gamma} \delta x^\gamma 
\right]_{,\mu\nu}}_{(*1)}
= \left[
    \frac{1}{2\kappa} e\, \eta^{ce} 
    \left( e^\nu_e e^\mu_b - e^\mu_e e^\nu_b \right) 
    \omega^b_{c\gamma} \delta x^\gamma
\right]_{,\mu\nu} = 0
\end{align*}

\noindent Term \((*2)\) can be rewritten using EoM\#4:
\begin{align*}
\underbracket[0.4pt][0pt]{\left(
        \frac{\partial \mathscr{L}_{GR}}{\partial \omega^b_{c\nu,\mu}}
    \right)_{,\nu} \omega^b_{c\gamma} \delta x^\gamma}_{(*2)}
    = \left(
    -\frac{\partial \mathscr{L}_{GR}}{\partial \omega^b_{c\mu,\nu}}
\right)_{,\nu} \omega^b_{c\gamma} \delta x^\gamma 
\underbracket{=}_{EoM\#4}
-\frac{\partial \mathscr{L}_{GR}}{\partial \omega^b_{c\mu}} \omega^b_{c\gamma} \delta x^\gamma 
= -\frac{\partial \mathscr{L}_{GR}}{\partial \omega^b_{c\nu,\mu}} 
\left(
    \omega^b_{d\gamma} \omega^d_{c\nu}
    - \omega^b_{d\nu} \omega^d_{c\gamma}
\right) \delta x^\gamma
\end{align*}

\noindent Gathering all terms, we obtain:
\begin{align*} 
\partial_\mu \mathcal{J}^\mu_{\delta x\text{-GR}} 
&= \partial_\mu \left[
    -\frac{\partial \mathscr{L}_{GR}}{\partial \omega^b_{c\nu,\mu}} \omega^b_{c\nu,\gamma} \delta x^\gamma  
    + \underbracket[0.4pt][0pt]{\frac{\partial \mathscr{L}_{GR}}{\partial \omega^b_{c\nu,\mu}} \omega^b_{c\gamma,\nu} \delta x^\gamma}_{(*3)}
    - \underbracket[0.4pt][0pt]{\frac{\partial \mathscr{L}_{GR}}{\partial \omega^b_{c\nu,\mu}} 
    \left( 
        \omega^b_{d\gamma} \omega^d_{c\nu} 
        - \omega^b_{d\nu} \omega^d_{c\gamma}
    \right) \delta x^\gamma}_{(*2)}
    + \delta^\mu_\gamma \mathscr{L}_{GR} \delta x^\gamma
\right] \\
&= \partial_\mu \left[
    \left(
        -\frac{\partial \mathscr{L}_{GR}}{\partial \omega^b_{c\nu,\mu}} 
        \left(
            \omega^b_{c\nu,\gamma} 
            - \omega^b_{c\gamma,\nu} 
            + \omega^b_{d\gamma} \omega^d_{c\nu} 
            - \omega^b_{d\nu} \omega^d_{c\gamma}
        \right) 
        + \delta^\mu_\gamma \mathscr{L}_{GR}
    \right) \delta x^\gamma
\right] \\
&= \partial_\mu \left[
    \underbracket[0.4pt][0pt]{
        \left(
            -\frac{\partial \mathscr{L}_{GR}}{\partial \omega^b_{c\nu,\mu}} 
            \mathscr{R}^b_{c\gamma\nu} 
            + \delta^\mu_\gamma \mathscr{L}_{GR}
        \right)
    }_{(t_{GR})^\mu_\gamma}
    \delta x^\gamma
\right]
\end{align*}

\noindent Therefore, the canonical energy-momentum tensor density for gravity reads:
\begin{align} 
\left(t_{GR}\right)^\mu_\gamma 
&= -\frac{\partial \mathscr{L}_{GR}}{\partial \omega^b_{c\nu,\mu}} \mathscr{R}^b_{c\gamma\nu} 
+ \delta^\mu_\gamma \mathscr{L}_{GR} \nonumber \\
&=- \frac{1}{2\kappa} e\, \eta^{ce} \left( e^\mu_e e^\nu_b - e^\nu_e e^\mu_b \right) \mathscr{R}^b_{c\gamma\nu} 
+ \delta^\mu_\gamma \mathscr{L}_{GR} \nonumber
\end{align}

\noindent Using the identity between spin curvature and Riemann curvature, this is 
\begin{align} 
\left(t_{GR}\right)^\mu_\gamma 
&= \frac{1}{2\kappa} 
\left(
    -g^{\beta\nu} R^\mu_{\beta\gamma\nu} 
    -g^{\alpha\mu} R_{\alpha\gamma} 
    + \delta^\mu_\gamma R 
\right) \sqrt{-g} \label{eq:Noether_t}
\end{align}
Furthermore, the well-known symmetry of the Riemann curvature tensor can be used, since the metric-compatible connection must be torsion-free indicated by EoM\#2, 
\begin{align*}
g^{\beta\nu} R^\mu_{\beta\gamma\nu}
= g^{\beta\nu} g^{\alpha\mu} R_{\alpha\beta\gamma\nu}
= g^{\beta\nu} g^{\alpha\mu} R_{\beta\alpha\nu\gamma}
= g^{\alpha\mu} R^\nu_{\alpha\nu\gamma}
= g^{\alpha\mu} R_{\alpha\gamma}
\end{align*}

\noindent Substituting into Eq.~\eqref{eq:Noether_t} gives:
\begin{align} \label{eq:Einstein=t}
\left(t_{GR}\right)^\mu_\gamma 
= \frac{1}{\kappa} 
\underbracket[0.4pt][0pt]{
    \left( 
        -g^{\alpha\mu} R_{\alpha\gamma} 
        + \frac{1}{2} \delta^\mu_\gamma R 
    \right)
}_{-g^{\mu\alpha} \mathbb{G}_{\alpha\gamma}} 
\sqrt{-g}
\end{align}

\noindent Thus, the canonical energy-momentum tensor density derived from Noether's theorem coincides with the Einstein tensor multiplied by the volume element \(\sqrt{-g}\). It satisfies the following key properties:
\begin{itemize}
    \item[\textbf{1.}] It is symmetric in indices \((\mu, \gamma)\).
    \item[\textbf{2.}] It is manifestly covariant: constructed solely from tensorial quantities.
    \item[\textbf{3.}] It vanishes in vacuum: \( R_{\mu\nu} = 0 \Rightarrow (t_{GR})^\mu_\gamma = 0 \), irrespective of coordinate or connection choices.
\end{itemize}

\section{Equivalence of Einstein-Hilbert and Noether Canonical Energy-Momentum Tensors}
\label{sec:Equivalence EMT}

\noindent In the previous sections, we derived the canonical energy-momentum tensor (EMT) using Noether's theorem. The result was a symmetric, gauge-invariant expression. In this section, we demonstrate that the Hilbert EMT—defined via the variation of the action with respect to the metric—and the canonical EMT are naturally equivalent. 

\medskip

\noindent The Lie derivative of the metric tensor \( g^{\gamma\alpha} \) with respect to the infinitesimal coordinate variation \( \delta x^\mu \) is given by:
\begin{align*}
\left( \hat{\mathcal{L}}_{\delta x} g \right)^{\gamma\alpha}
= -g^{\mu\alpha} {\delta x^\gamma}_{;\mu} - g^{\gamma\mu} {\delta x^\alpha}_{;\mu}
\end{align*}

\noindent Consider the action with Lagrangian density \( \mathscr{L} = \mathscr{L}[g^{\gamma\mu}, A_\nu, A_{\nu,\mu}, x^\gamma] \). Its total variation reads:
\begin{align*}
\Delta S &=
\int \frac{\delta \mathscr{L}}{\delta g^{\gamma\alpha}}\, \delta g^{\gamma\alpha} \, d^4x 
+ \int \cancelto{0}{\left[\text{EoM of } \delta A_\nu \right]} \delta A_\nu \, d^4x 
+ \int \partial_\mu \left[  
    \frac{\partial \mathscr{L}}{\partial(\partial_\mu A_\nu)} \Delta A_\nu
    + \underbracket[0.4pt][0pt]{\left( 
        -\frac{\partial \mathscr{L}}{\partial(\partial_\mu A_\nu)} F_{\gamma\nu} 
        + \delta^\mu_\gamma \mathscr{L} 
    \right)}_{t^\mu_\gamma} \delta x^\gamma  
\right] d^4x
\end{align*}

\noindent To study symmetry under spacetime translations \( \delta x^\mu \), we set \( \delta g = -\hat{\mathcal{L}}_{\delta x} g \) and \( \delta A = -\hat{\mathcal{L}}_{\delta x} A \Rightarrow \Delta A = 0 \). The variation becomes:
\begin{align*}
\Delta S &=
\int -\frac{\delta \mathscr{L}}{\delta g^{\gamma\alpha}} \left( \hat{\mathcal{L}}_{\delta x} g \right)^{\gamma\alpha} \, d^4x 
+ \int \partial_\mu \left[ t^\mu_\gamma \delta x^\gamma \right] d^4x \\
&=
\int 2 \frac{\delta \mathscr{L}}{\delta g^{\gamma\alpha}} g^{\mu\alpha} {\delta x^\gamma}_{;\mu} \, d^4x
+ \int \partial_\mu \left( \sqrt{-g} \, T^\mu_\gamma \delta x^\gamma \right) d^4x \\
&=
\int 2 \frac{\delta \mathscr{L}}{\delta g^{\gamma\alpha}} g^{\mu\alpha} {\delta x^\gamma}_{;\mu} \, d^4x
+ \int \sqrt{-g} \, \nabla_\mu \left( T^\mu_\gamma \delta x^\gamma \right) d^4x \\
&=
\int 2 \frac{\delta \mathscr{L}}{\delta g^{\gamma\alpha}} g^{\mu\alpha} {\delta x^\gamma}_{;\mu} \, d^4x
+ \int \sqrt{-g} \left( \nabla_\mu T^\mu_\gamma \right) \delta x^\gamma \, d^4x
+ \int \sqrt{-g} \, T^\mu_\gamma {\delta x^\gamma}_{;\mu} \, d^4x
\end{align*}
. Here \(\nabla_\mu\) is the Levi-Civita covariant derivative.
\noindent The first and the third terms combine into:
\begin{align*}
\Delta S =
\int \left[
    2 \frac{\delta \mathscr{L}}{\delta g^{\gamma\alpha}} g^{\mu\alpha}
    + \sqrt{-g} \, T^\mu_\gamma
\right] {\delta x^\gamma}_{;\mu} \, d^4x
+ \int \sqrt{-g} \left( \nabla_\mu T^\mu_\gamma \right) \delta x^\gamma \, d^4x
\end{align*}

\noindent If the action is invariant under translations, then \( \Delta S = 0 \), then $\nabla_\mu T^\mu_\gamma=0$ is the conservation law of energy momentum in the curved spacetime. This leads to the first term must be 0:

\begin{align*}
\underbracket[0.4pt][0pt]{T^\mu_\gamma}_{\text{Canonical EMT}}
= g^{\mu\alpha} 
\underbracket[0.4pt][0pt]{\left( -\frac{2}{\sqrt{-g}} \frac{\delta \mathscr{L}}{\delta g^{\gamma\alpha}} \right)}_{\text{Hilbert EMT}}
\end{align*}

\noindent This identity shows that the canonical energy-momentum tensor derived via Noether’s theorem is equivalent to the Hilbert EMT defined via metric variation. This demonstrates their natural equivalence within a diffeomorphism-invariant Lagrangian field theory.

\section{Alternative Derivation of the Einstein Field Equations}
\label{sec:Einstein eq}

\noindent Since the canonical energy-momentum tensor (EMT) and the Hilbert EMT are equivalent, we may derive the Einstein field equations directly from Noether’s theorem, without resorting to metric variation. Consider the total action composed of gravitational and electromagnetic contributions:
\begin{align*} 
S = \int d^4x \left( \mathscr{L}_{GR} + \mathscr{L}_{EM} \right)
\end{align*}

\noindent The total variation under infinitesimal spacetime translation \( x^\gamma \rightarrow x^\gamma + \delta x^\gamma \) induces the following field transformations:
\begin{align*} 
A_\nu &\rightarrow A_\nu + \underbracket[0.4pt][0pt]{\delta A_\nu + \left(\hat{\mathcal{L}}_{\delta x} A\right)_\nu}_{\Delta A_\nu} \\
\omega^b_{c\nu} &\rightarrow \omega^b_{c\nu} + \underbracket[0.4pt][0pt]{\delta \omega^b_{c\nu} + \left(\hat{\mathcal{L}}_{\delta x} \omega\right)^b_{c\nu}}_{\Delta \omega^b_{c\nu}}
\end{align*}

\noindent Plugging these into the variation of the action yields:
\begin{adjustwidth}{-2.3cm}{-1cm}
\begin{align*} 
\Delta S = \int d^4x \Bigg\{ 
& \left[ 
    \frac{\partial \mathscr{L}_{GR}}{\partial \omega^b_{c\nu}} 
    - \left( \frac{\partial \mathscr{L}_{GR}}{\partial \omega^b_{c\nu,\mu}} \right)_{,\mu} 
\right] \delta \omega^b_{c\nu}
+ \left[ 
    \frac{\partial \mathscr{L}_{EM}}{\partial A_\nu} 
    - \left( \frac{\partial \mathscr{L}_{EM}}{\partial A_{\nu,\mu}} \right)_{,\mu} 
\right] \delta A_\nu \\
& + \partial_\mu \left[ 
    \frac{\partial \mathscr{L}_{GR}}{\partial \omega^b_{c\nu,\mu}} \Delta \omega^b_{c\nu} 
    + \frac{\partial \mathscr{L}_{EM}}{\partial A_{\nu,\mu}} \Delta A_\nu 
    + \left( (t_{GR})^\mu_\gamma + (t_{EM})^\mu_\gamma \right) \delta x^\gamma 
\right] 
\Bigg\}
\end{align*}
\end{adjustwidth}

\noindent The last surface term encodes spacetime symmetry under translations. Using the known identity for divergence in curved space, we write:
\begin{align*} 
\partial_\mu \left\{\left[ \left( t_{GR} \right)^\mu_\gamma + \left( t_{EM} \right)^\mu_\gamma \right] \delta x^\gamma \right\} = 0 
\quad \Rightarrow \quad 
\sqrt{-g}\,\nabla_\mu \left\{\left[ \left( T_{GR} \right)^\mu_\gamma + \left( T_{EM} \right)^\mu_\gamma \right] \delta x^\gamma \right\}  = 0
\end{align*}

\noindent If the translation is global, meaning \( \nabla_\mu \delta x^\gamma = 0 \), then the conservation law reduces to:
\begin{align*} 
\nabla_\mu \left[ \left( T_{GR} \right)^\mu_\gamma + \left( T_{EM} \right)^\mu_\gamma \right] = 0
\end{align*}

\noindent Therefore, the total energy-momentum tensor must be conserved. By the structure of conserved symmetric tensors in general relativity, we have:
\begin{align*} 
\left( T_{GR} \right)^\mu_\gamma + \left( T_{EM} \right)^\mu_\gamma = \delta^\mu_\gamma \Lambda
\end{align*}
where \( \Lambda \) is an integration constant, interpreted as the cosmological constant. Substituting the explicit form of the gravitational energy-momentum tensor from Eq.~\eqref{eq:Einstein=t}, we obtain:
\begin{align*} 
\mathbb{G}_{\alpha\gamma} + g_{\alpha\gamma} \Lambda = \kappa g_{\mu\alpha} \left( T_{EM} \right)^\mu_\gamma
\end{align*}

\noindent This is the Einstein field equation with cosmological constant, derived purely from Noether’s theorem applied to translation symmetry, without invoking the variation of the metric tensor.

\section{Summary}

\noindent
In this work, we systematically explored the derivation of the energy-momentum tensor (EMT) from Noether’s theorem within both gauge theory and general relativity. The analysis began by emphasizing that the gauge potential should be treated as a 1-form field rather than a vector field. This differential geometric perspective ensures that the total variation under spacetime translations correctly produces a symmetric, gauge-invariant canonical EMT—without requiring any artificial symmetrization. This construction remains valid on arbitrary curved manifolds.

We then extended this framework to the gravitational case using the vielbein formalism. In this approach, the spin connection plays the role of a gauge field valued in the local Lorentz group, and the Hilbert–Einstein action becomes analogous to a gauge-theoretic Lagrangian. Applying Noether’s theorem to local spacetime transformations, we derived a canonical energy-momentum tensor that matches, up to a factor of the volume form, the Einstein tensor. This derivation demonstrates that the tensor is not only symmetric and generally covariant, but also vanishes in vacuum, making it physically meaningful.

The equivalence between the canonical EMT derived via Noether’s theorem and the Hilbert EMT obtained through metric variation was then established explicitly. This result highlights that energy-momentum conservation in curved spacetime naturally arises from the variational principle associated with coordinate invariance.

Finally, we presented an alternative route to the Einstein field equations that avoids varying the metric altogether. By considering the total variation of the full action under spacetime translation symmetry and requiring the conservation of the total Noether current, we recovered the Einstein equation with a cosmological constant term.

In conclusion, this unified treatment demonstrates how Noether’s theorem serves as a principle for constructing and interpreting the dynamics of gauge fields and spacetime geometry on equal footing.

\section*{Acknowledgments}

I would like to thank Yen-Cheng Chang, Dean-Yi Chou for helpful discussions. In memory of T.-Y. Liu.

\bibliography{references.bib}

\end{document}